\documentstyle[aps,12pt,manuscript]{revtex}
\begin{document}
\preprint{INJE--TP--96--6}
\def\overlay#1#2{\setbox0=\hbox{#1}\setbox1=\hbox to \wd0{\hss #2\hss}#1%
\hskip -2\wd0\copy1}

\title{Classical instanton and wormhole solutions of Type IIB string theory }

\author{Jin Young Kim$^1$, H.W. Lee$^2$, and Y.S. Myung$^2$}
\address{$^1$ Division of Basic Science, Dongseo University, Pusan 616--010, Korea\\
$^2$Department of Physics, Inje University, Kimhae 621--749, Korea} 

\maketitle

\vskip 1.5in

\begin{abstract}
We study $p=-1$ D-brane in type IIB superstring theory.
In addition to RR instanton, we obtain the RR charged wormhole solution in the Einstein frame.
This corresponds to the ten-dimensional  singular wormhole solution with infinite
euclidean action.
\end{abstract}

\newpage
There has been much progress in nonperturbative aspects of string theory\cite{Duf}.
Recently some $p$-brane solitons appeared in the form of solitonic 
objects(Dirichlet(D)-branes) as well as D-instantons\cite{Pol}.
The $p$ D-branes were used mainly to understand the microscopic origin of
the black hole entropy\cite{Vafa}. In addition to solitonic objects, 
type IIB superstring theory possesses
an instanton  solution in ten dimensions\cite{Gib,Gub}. This is  the $p=-1$ D-brane
 which carries 
the Ramond-Ramond (RR) charge. Such instantons are important in the theory. For example,
they are responsible for point-like effects in fixed-angle scattering. We remind the reader that 
the possibility of inducing point-like structure is one of the original 
reasons for investigating string theories 
with Dirichlet boundary conditions\cite{Gre}.

On the other hand, there are many kinds of euclidean wormhole solutions.
In four-dimensions the following matter fields which support the throat of the wormhole were adopted:
axion fields\cite{Gid1}, scalar fields\cite{Lee}, SU(2) Yang-Mills fields\cite{Hos}.
 Higher-dimensional wormhole solutions
were obtained \cite{Yos,Mye} and higher-derivative correction to the Einstein-Hilbert
 action was considered\cite{Fuk}.
Especially, it is shown that there is
no-nonsigular wormhole solution for the axion in four-dimensional stringy model which includes
 graviton, dilaton and axion\cite{Gid1,Gid2}. This is because 
the dilation is nontrivially coupled to metric and axion. 
The axion (pseudoscalar) is dual of  the NS-NS three-form $H=dB$
and plays the role of the source for the wormhole.

In this paper we will study $p=-1$ D-brane of type IIB superstring  in the Einstein frame.
We obtain, in addition to RR instanton (classical D-instanton), 
the new RR charged wormhole solution in ten-dimensional space
with the euclidean signature $(++ \cdots)$.
We are interested in the contribution of RR instanton and RR charged wormhole configurations to
 the euclidean functional integral
for the forward ``flat space $\to$ flat space '' amplitude\cite{Rub}.
Wormholes$-$solutions to the euclidean Einstein equations that connect two asymptotically
flat regions$-$are considered as saddle points of this integral and
are very important for semiclassical calculations of transition probabilities of
 topological
change  in quantum gravity.
 If the action of wormhole is given by $(S)$,
the transition probability of topological change is proportional to $e^{-S}$. 
For the  wormhole of metric-axion system\cite{Gid1},
 $S$ is positive and large for large axionic
charge flowing through the wormhole. And then the transition probability is small and the wormhole
transition can be  suppressed. On the other hand, for the stringy wormhole of dilaton-metric-axion
the transition probability is almost zero because $S$ is infinite\cite{Gid2}.

We start with  the  ten-dimensional form of type II superstring action\cite{Gibb,Hor}, 

\begin{equation}
S_S = \int d^{10} x \sqrt{-g_S} 
  \Big[ e^{-2\phi}( R_S  + 4 (\nabla \phi)^2) - { 1 \over 2(p+2)!} F^2_{p+2} \Big],
\end{equation}
where $F_{p+2}$ is an RR (p+2)-form field strength $(F_{p+2}=dA_{p+1})$
 and the subscript $S$ indicates that
the string metric is being used.
Here we observe that kinetic term of RR field is not multiplied by the 
dilaton factor ($e^{-2 \phi}$)\cite{Wit}. This is because the special coupling of the RR fields
to the dilaton lies in the structure of local $N=2, D=10$ supersymmetry. 
In the Einstein frame with the signature $(-+\cdots)$,
through the relation $g_{S~MN}=e^{\phi/2} g_{MN}$, one finds the action as
\begin{equation}
S = \int d^{10} x \sqrt{-g} 
  \Big[ R  - {1 \over 2} (\nabla \phi)^2 - {1 \over 2 (p+2)!}e^{-(p-3) \phi/2} F^2_{p+2} \Big].
\end{equation}
The even (odd) $p$ correspond to type IIA (type IIB) superstrings. 
The  solution of (2) for all values from $p=-1$ to $p=9$ was found in\cite{Gibb,Hor}.
 In this paper we consider
only the case of $p=-1$.
 This is  a special case because 
it requires a euclidean continuation.  $A_{0}$-form is just an RR scalar ($a$) which 
is a source of the RR instanton and RR charged wormhole. 
This  is essentially a ten-dimensional axion.
First, we review the RR instanton\cite{Gib}. The corrsponding action is given by 
\begin{equation}
S_{p=-1} = \int d^{10} x \sqrt{-g} 
  \Big[ R  - {1 \over 2} (\nabla \phi)^2 - {1 \over 2 }e^{2\phi}(\nabla a)^2  \Big].
\end{equation}
Defining a nine-form field strength as $F_9=e^{2 \phi} *da$, the action (3) can be rewritten in 
the equivalent dual-form,
\begin{equation}
S_{p=7} = \int d^{10} x \sqrt{-g} 
  \Big[ R  - {1 \over 2} (\nabla \phi)^2 - {1 \over 2 (9!)}e^{-2 \phi} F^2_9 \Big].
\end{equation}
The equations of motion for the euclidean version can be derived from (3) with the substitution
$a \to \alpha= ia$. These are 
\begin{eqnarray}
&R_{MN} -{1 \over 2}( \nabla_M \phi \nabla_N \phi- e^{2 \phi}
    \nabla_M \alpha \nabla_N \alpha) &=0, \nonumber   \\
& \nabla_M ( e^{2 \phi} \partial^M \alpha) &= 0,   \\
& \nabla^2 \phi + e^{2 \phi} (\nabla \alpha)^2 &=0.   \nonumber
\end{eqnarray}
We consider the conditions that need to be satisfied for a solution of 
the euclidean theory to preserve half of the $N=2$ supersymmetry of the tpye IIB theory.
These are
\begin{equation}
d \alpha =  e^{-\phi} d \phi, ~~~~ ds^2= d\rho^2 + \rho^2 d\Omega^2_9,
\end{equation}
where $\rho$ is the radius of nine sphere ($S^9$) and its line element ($d \Omega^2_9$).
Substitution of these into (5) leads to $\partial^2 \phi= -(\partial \phi)^2$ and 
thus $\partial^2 (e^{\phi})=0$. This allows us a spherically symmetric solution which
describes a single RR instanton,
\begin{equation}
e^{\phi} = e^{\phi_{\infty}} + {c \over \rho^8},
\end{equation}
where $\phi_{\infty}$ is the value of the dilaton field at $\rho=\infty$ and $c$ is a constant
that will be shown later to be proportional to the instanton charge. 
This instanton solution is evidently singular at $\rho=0$ in the Einstein frame. 
However, in the string frame, one finds the metric

\begin{equation}
 ds_S^2= e^{\phi/2} ds^2 =\sqrt { e^{\phi_{\infty}} + {c \over \rho^8}}
( d\rho^2 + \rho^2 d\Omega^2_9).
\end{equation}
The above metric is invariant under the inversion transformation 

\begin{equation}
\rho \to  (c e^{-\phi_{ \infty}})^{1/4} {1 \over \rho}. 
\end{equation}
This shows that the region of $\rho \to 0$ is another asymptotically euclidean region
which is identical to that of $\rho \to \infty$. The solution in the string frame is thus a
wormhole in which there exist two asymptotically euclidean regions connected by a neck.

Now we wish to find the RR charged wormhole solution in the Einstein frame.
The euclidean action for this purpose is given by 
\begin{equation}
S_{worm} = \int_M  
  \big[ - R  + {1 \over 2} (\nabla \phi)^2 + {1 \over 2 }e^{2\phi}(\nabla a)^2  \big]
 -2 \int_{\partial M} [ {\rm Tr} K ],
\end{equation}
where $M (\partial M =S^9)$ is a ten-dimensional euclidean space (its boundary) 
and [Tr$ K$] is  the boundary contribution.
 However, the boundary contribution is not  relevant here.
We note here that the RR scalar ($a$) can be considered as the source for
 the RR instanton and RR charged wormhole\cite{Gid2}.
One thus takes the Noether current $J_M =e^{2 \phi}\partial_M a$ and requires its conservation
\begin{equation}
\partial_M (\sqrt {g} J^M) =0
\end{equation}
which is equivalent to the field equation in (5).
Therefore we have to perform the functional integration over conserved current densities.   
Let us introduce the general $O(10)$-symmetric euclidean metric as

\begin{equation}
 ds^2= N^2(\rho) d\rho^2 + R^2(\rho) d\Omega^2_9
\end{equation}
with two scale factors $(N, R)$.
The $O(10)$-symmetric curent density has one non-zero component ($J^0(\rho))$ and its
conservation in (11) means that $\sqrt g J^0$ is a constant. This constant is related to the 
global charge $Q$ of the RR instanton $({Q / Vol(S^9)})$. Thus one finds

\begin{equation}
 J^0 = {12 Q \over \pi^5} {1\over N R^9}.
\end{equation}
Then the $O(10)$-symmetric action is given by
\begin{equation}
S_{worm}= - 6 \pi^5 \int d\rho \big[ { R^7 \over N} (\partial_\rho R)^2 + N R^7
-{1 \over 144}{R^9 \over N} (\partial_\rho \phi)^2 -{Q^2 \over \pi^{10}}
{N \over R^9} e^{-2 \phi} \big].
\end{equation}
From the above action, we find three equations
\begin{equation}
-{ R^7 \over N^2} (\partial_\rho R)^2 +  R^7
+{1 \over 144}{R^9 \over N^2} (\partial_\rho \phi)^2 -({Q \over \pi^5})^2
{ e^{-2 \phi} \over R^9} =0,  
\end{equation}

\begin{equation}
 \nabla_\rho( {R^9 \partial_\rho \phi \over N^2}) + 
 ({12 Q \over \pi^5})^2{ N e^{-2 \phi} \over R^9} = 0,  
\end{equation}

\begin{equation}
 7 { R^6 \over N} (\partial_\rho R)^2 - 2 \partial_\rho( {R^7 \partial_\rho R \over N}) + 7 N R^6
-{9 \over 144}{R^8 \over N} (\partial_\rho \phi)^2 + 9 ({Q \over \pi^5})^2
{ N e^{-2 \phi} \over R^{10}} =0.   
\end{equation}
For simplicity we choose the  $N=1$ gauge.
We  then  obtain the exact solution of (15)-(17),
\begin{equation}
N^I=1,~~~ (R^I)^2= \rho^2,~~~~ e^{\phi^I} = e^{\phi^I_\infty} + 
{3 \over 2}{Q \over \pi^5} {1 \over \rho^8}.
\end{equation}
Comparing (18) with (6) and (7), one immediately recovers 
the RR instanton solution with $c= {3 Q /2 \pi^5}$.

With different boundary conditions, we can also find the new
 wormhole (RR charged wormhole) solution.
First, let us consider the region of $\rho=0$.
 From (16) one obtains
the first integral equation for the dilaton
\begin{equation}
(\partial_\rho \phi)^2 = 
 \big({12 Q \over \pi^5} \big)^2 \big( {e^{-2 \phi} \over R^{18}} - 
{e^{- 2 \phi^W_0} \over R^{18}} \big).
\end{equation}
Here the integration constant has been fixed so that $\phi$ has vanishing derivative at $\rho=0$.
This point will be  the RR charged wormhole neck with the radius $R^W(\rho=0) =R_0$.
At this point one requires $\partial_\rho R^W|_{\rho=0}=0$ and
$R_0^{16}= (Q/\pi^5)^2 e^{-2 \phi^W_0}$.
(19) is derived actually by integration of (16) from 0 to $\rho$.  
Substituting (19) into the $(\rho\rho)$-metric equation (15) leads to the important equation
\begin{equation}
(\partial_\rho R)^2 = 1 - {R_0^{16} \over R^{16}}.
\end{equation}
Further, using (19) and (20), it turns out that the angular-component metric equation (17)
is redundant and leads to (20).
In order to study the behavior of wormhole dilaton, let us
solve (19) to obtain
\begin{equation}
e^{2 \phi} = ({Q \over \pi^5 R_0^8})^2 \cos^2 \big[ {3 \over 2} \cos^{-1} ({R_0 \over R})^8 \big].
\end{equation}
When $R^W(\rho_{cr})=2^{1/8} R_0$ for finite $\rho= \rho_{cr}$, the right-hand side 
is zero and thus one has $\phi^W(\rho_{cr}) = - \infty$.
 Substituting this into (14), the last term (action density) gives rise to an infinity at 
$\rho=\rho_{cr}$. It suppresses the transition probability $(e^{-S_{worm}})$ completely.
When  a dilaton is coupled to metric and axion as in (3), one can always find singular solution
with infinite euclidean action. This shows there is no nonsingular RR charged wormhole solution
for $p=-1$ D-brane of type IIB theory.

In order to obtain the large $\rho$-behavior $(\rho > \rho_{cr})$, 
one has to integrate (16) from $\infty$ to $\rho$.  In this case, one also finds the same 
forms as in (19), (20) and (21) but with the substitutions:
 $\phi^W_0 \to \phi^W_\infty + (1/2)\ln 2; R^{16}_0 \to R^{16}_\infty =
{1 \over 2}(Q/\pi^5)^2 e^{-2 \phi^W_\infty}$. These are given by
\begin{equation}
(\partial_\rho \phi)^2 = 
 \big({12 Q \over \pi^5} \big)^2 \big( {e^{-2 \phi} \over R^{18}} - 
{e^{- 2 \phi^W_\infty} \over 2 R^{18}} \big),
\end{equation}

\begin{equation}
(\partial_\rho R)^2 = 1 - {R_\infty^{16} \over R^{16}},
\end{equation}

\begin{equation}
e^{2 \phi} = ({Q \over \pi^5 R_\infty^8})^2 \cos^2 \big[ {3 \over 2} \cos^{-1} 
({R_\infty \over R})^8 \big].
\end{equation}
Now we consider the asymptotically flat region. 
From (23) the resulting solution  has asymptotic behavior
 $R^W(\rho) \to \pm \rho$ for $\rho \to  \pm \infty$. On the other hand,  it  has minimum at 
the wormhole neck ($R_0$). 
This corresponds to  ten-dimensional RR charged wormhole solution, which is based on the RR sector.
Here we wish to point out the distinction between RR instanton and RR charged wormhole.  
 This is related to the different behavior
of the dilation at the boundary ($\rho=0$). In the case of RR instanton, 
one has $\phi^I(0) \to \infty$ and
$\phi^I(\infty) \to \phi^I_\infty$= constant. On the other hand,
although the asymptotic structure of the RR charged wormhole is the same as that of the 
RR instanton, 
we have  $\phi^W(0)=\phi^W_0 (=\exp\{ (Q/\pi^5)^2 R_0^{-16} \})$ with 
$\partial_\rho \phi^W |_{\rho=0}=0$. That is, we have the singular solution of the dilaton 
at $\rho=0$  for the RR instanton, while the non-singular solution at $\rho=0$
is required for the RR charged wormhole.

For an explicit calculation, we first have to solve the differential equations (20) and (23)
 by numerical analysis.
  We introduce the rescalings
 $(\rho/\rho_0, R/R_0, \phi/\phi^W_0)$ with $\rho_0=R_0$ for (20), (21) and
$(\rho/\rho_0, R/R_\infty, \phi/\phi^W_\infty)$ with $\rho_0=R_\infty$ for (23), (24). 
 The resulting solution 
is shown in Fig.1 for $R_0=R_\infty$ and $\phi^W_0=\phi^W_\infty$. 
  Far from the RR charged wormhole throat $(\rho/R_0 > 1)$, 
one can ignore the effect of gravity
and the euclidean space becomes flat $(R^W \sim \rho)$.  Here one can find the wormhole neck
($\partial_\rho R^W =0$) near $\rho = 0$.  
Also one can get the point 
$\rho = \rho_{cr}$,  as a solution to the relation $R^W(\rho_{cr}) = 2^{0.125} R_0$. 
Now let us substitute the result of $R^W(\rho)/R_0$ in Fig.1 into (21) and (24).  
Then one obtains the wormhole dilaton ($\phi^W(\rho)$)-behavior.
As is shown in Fig.2, $\phi^W$ is obviously singular at $\rho = \rho_{cr} \simeq 1.08 R_0$.
For $\rho < \rho_{cr}$, one can observe the feature of RR charged wormhole throat. 
 Clearly $\phi^W$ is different from $\phi^I$. 
The instanton dilation ($\phi^I$) is  singular
at $\rho=0$, while the wormhole dilaton ($\phi^W$) is non-singular at $\rho=0$.
 On the other hand, for   $\rho > \rho_{cr}$, $\phi^W$ approaches 
 $\phi^I$. 

In conclusion, we find the RR charged wormhole solution in $p=-1$ D-brane of type IIB 
superstring theory.
Unfortunately, this is a singular solution with an infinite euclidean action.
This is because kinetic term of RR field is not multiplied by the 
dilaton factor ($e^{-2 \phi}$) in the string frame.
Also we clarify the distinction bewteen the known RR instanton and RR charged wormhole solutions.
This is mainly due to the different dilaton configurations.
Finally we point out that the RR instanton is supersymmetric, while the RR charged wormhole
is not supersymmetric. This is because the the half of $N=2$ supersymmetry is preserved when
the metric is flat as in the case of RR instanton (18). For this case 
the solution is easily found and thus the numerical integration is not necessary.
  It is obvious that the RR charged wormhole
do not preserve any supersymmetry because its metric is not flat as in (20).

\acknowledgments

This work was supported in part by the Basic Science Research Institute 
Program, Ministry of Education, Project NOs. BSRI--96--2413, BSRI--96--2441
and by Inje Research and Scholarship Foundation.

\figure{Fig. 1: $R/R_0$ as a function of $\rho / R_0$.  The solid, dashed and dotted
lines correspond to  $R/R_0 = 2^{0.125}$,  RR charged wormhole scale factor ($R^W/R_0$) and 
 RR instanton scale factor ($R^I/R_0$) respectively. 
The singular critical point ($\rho_{cr}$) is determined from the solution of
 $R^W(\rho_{cr})/R_0 = 2^{0.125}$.

Fig. 2: $\phi / \phi_0$ as a function of $\rho / R_0$.  The solid and dashed 
lines correspond to RR charged wormhole dilaton ($\phi^W/\phi^W_0$) and RR instanton dilaton 
($\phi^I / \phi^W_0$) respectively. 
The singular point  is found at $\rho_{cr} \simeq 1.08 R_0$.}

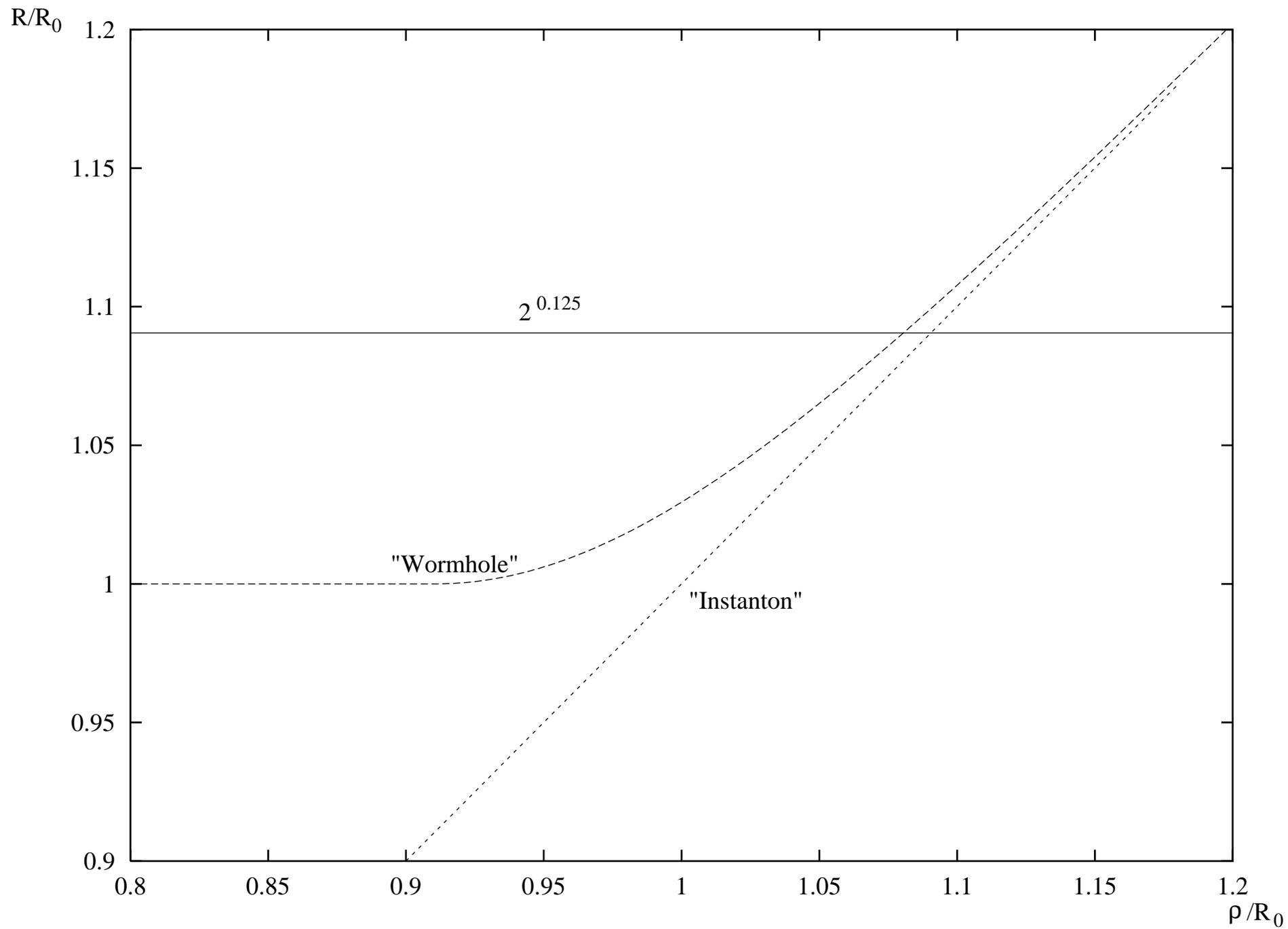
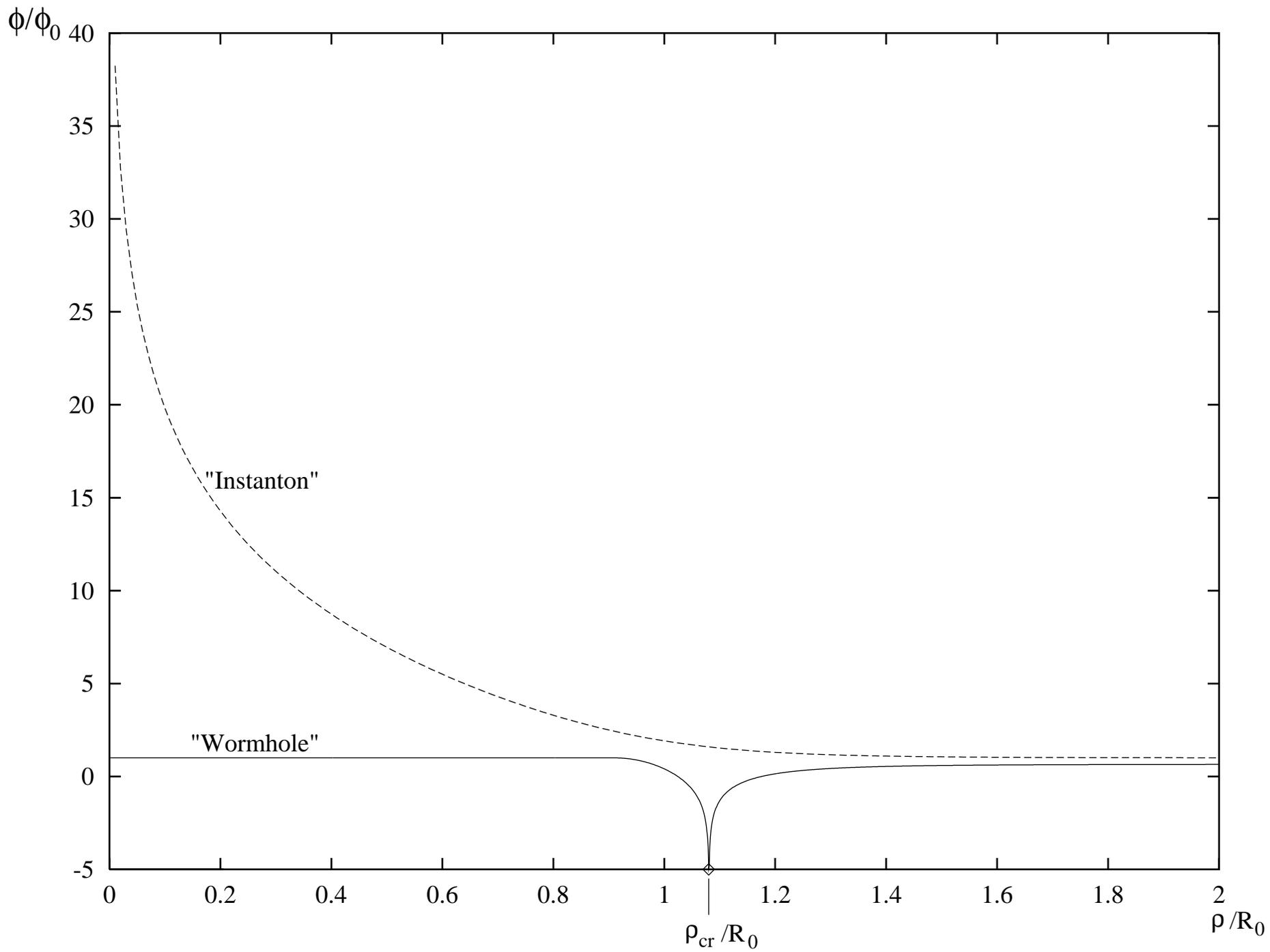
\end{document}